\documentclass[conference,letterpaper]{IEEEtran}

\usepackage[utf8]{inputenc}
\usepackage[T1]{fontenc}
\usepackage{url}
\usepackage{ifthen}
\usepackage{cite}
\usepackage{graphicx}
\usepackage{amsthm}
\usepackage[cmex10]{amsmath} 
\usepackage[linesnumbered,ruled,vlined]{algorithm2e}
\usepackage{fancyhdr}

\begin{document}
\title{Semantic Arithmetic Coding using Synonymous Mappings}

\author{%
    \IEEEauthorblockN
    {
        Zijian~Liang\IEEEauthorrefmark{1},
        Kai~Niu\IEEEauthorrefmark{2},
        Jin~Xu\IEEEauthorrefmark{1},
        and Ping~Zhang\IEEEauthorrefmark{2}
    }
    \IEEEauthorblockA
    {\IEEEauthorrefmark{1}%
        The Key Laboratory of Universal Wireless Communications, Ministry of Education
    }
    \IEEEauthorblockA
    {\IEEEauthorrefmark{2}%
        The State Key Laboratory of Networking and Switching Technology
    }
    \IEEEauthorblockA
    {
        Beijing University of Posts and Telecommunications, Beijing 100876, China \\
        Email: \{liang1060279345, niukai, xujinbupt, pzhang\}@bupt.edu.cn
    }
}
\maketitle
\thispagestyle{fancy}


\begin{abstract}
    Recent semantic communication methods explore effective ways to expand the communication paradigm and improve the system performance of the communication systems. Nonetheless, the common problem of these methods is that the essence of semantics is not explicitly pointed out and directly utilized. A new epistemology suggests that synonymy, which is revealed as the fundamental feature of semantics, guides the establishment of the semantic information theory from a novel viewpoint. Building on this theoretical basis, this paper proposes a semantic arithmetic coding (SAC) method for semantic lossless compression using intuitive semantic synonymy. By constructing reasonable synonymous mappings and performing arithmetic coding procedures over synonymous sets, SAC can achieve higher compression efficiency for meaning-contained source sequences at the semantic level and thereby approximate the semantic entropy limits. Experimental results on edge texture map compression show an evident improvement in coding efficiency using SAC without semantic losses, compared to traditional arithmetic coding, which demonstrates its effectiveness.
\end{abstract}

\section{Introduction}

In recent years, research on semantic communications has taken a different development route from traditional communication technologies. While traditional communications separately optimize the source compression and data transmission guided by Shannon’s classical information theory (CIT) \cite{shannon1948mathematical, cover1999elements}, recent works on semantic communications mostly explore end-to-end performance optimization utilizing deep neural networks and joint source-channel coding frameworks. They expand semantic communication paradigms \cite{ping2022intellicise, niu2022paradigm, dai2022communication} and effectively improve the end-to-end performance of the communication systems for diverse source modalities oriented towards both point-to-point \cite{farsad2018deep, xie2021deep, weng2021semantic, bourtsoulatze2019deep, dai2022nonlinear, wang2022wireless} and multi-user transmission scenarios \cite{zhang2023model, zhang2023deepma, niu2023semantic}.

However, these works are too slavish, lacking reliance on appropriate semantic information theory. Although researchers have explored semantic information theory from various viewpoints ranging from the perspective of logical probability \cite{carnap1952outline, bar1953semantic, floridi2004outline, bao2011towards} to fuzzy information theory \cite{de1972definition, de1974entropy, al2001fuzzy} since Weaver discussed three-level communication problems \cite{weaver1953recent}, none of these theories can become a universal guiding theory for semantic communication methods. Furthermore, the absence of theoretical limits in semantic coding leads current semantic communication methods to utilize indicators at the syntactic level as optimization directions, such as mean-squared error \cite{bourtsoulatze2019deep} or Kullback-Leibler divergence \cite{dai2022nonlinear}. In these cases, a common problem of existing semantic communication methods is that the essence of semantics is not explicitly pointed out and directly utilized, which makes it unclear enough to determine whether a semantic coding method is semantically lossless.

In light of this, we delve deep into the meaning of semantics and propose a new epistemology for semantic information theory. That is, synonymy is the fundamental feature of semantic information, and synonymous mappings indicate the relationship between semantic information and syntactic information. Based on this novel viewpoint, a mathematical framework of semantic information theory is established in \cite{niu2024Mathematical}. As an important content, the semantic source coding theorem and its corresponding compression limit (i.e., semantic entropy) are determined. These theories reveal the methodology that by introducing synonymous sets and performing semantic compression over them, source compression efficiency can be further improved without semantic losses. We noticed that similar ideas have appeared in semantic compression methods for tabular data \cite{jagadish1999semantic, babu2001spartan, jagadish2004itcompress}; however, we should point out that this methodology should be adaptable to any source data type under the premise of well-designed synonymous mappings.

In this paper, we propose an arithmetic coding (AC) method for semantic lossless compression based on these theories, named semantic arithmetic coding (SAC). By constructing reasonable synonymous mappings to partition synonymous sets and performing the arithmetic encoding procedures over the synonymous set corresponding to the coding syntactic symbol, SAC can achieve higher compression efficiency under semantic lossless conditions. Moreover, the theoretical limit approachability to semantic entropy of our proposed SAC is validated based on an extension of the code length theorem of arithmetic codes and relative experimental verification.

\section{System Model and Theoretical Limits}

In this section, we briefly review the system model of semantic source coding, along with its theoretical compression limit based on the critical feature of synonymous mappings.

\subsection{System Model}

Semantic source coding is an extension of classic source coding under the guidance of semantic information theory, with its goal still being to compress source data. However, unlike classic source coding, semantic source coding focuses on ensuring no distinctions in implicit meanings between sequences before encoding and after decoding, without strictly requiring complete consistency in their explicit syntactical forms.

As stated in \cite{niu2024Mathematical}, all perceptible messages are syntactic information, and all such syntactic information is presented to illuminate the underlying semantic information. Therefore, we can establish the system model for semantic source coding as follows:
\begin{equation}\label{system_model}
    \tilde{u}^m \xrightarrow{f\left(\cdot\right)}  {u}^m \xrightarrow{e\left(\cdot\right)}  {b}^l \xrightarrow{d\left(\cdot\right)} \hat{u}^m \xrightarrow{g\left(\cdot\right)} \hat{\tilde{u}}^m,
\end{equation}
where $\tilde{u}^m$ and $\hat{\tilde{u}}^m$ are invisible source and reconstructed semantic variable sequences, and ${u}^m$ and $\hat{u}^m$ are perceptible source and reconstructed syntactic variable sequences, respectively, in which $m$ denotes the length of the source sequences. The mapping $f\left(\cdot\right)$ and its reverse $g\left(\cdot\right)$ represent the invisible conversion relationship between the semantic information and the syntactic information.

For the main process of the coding, the semantic source encoder $e\left(\cdot\right)$ operates on the syntactic sequence ${u}^m$, encoding it into a codeword sequence $ {b}^l$ of length $l$, and the corresponding semantic source decoder $d\left(\cdot\right)$ transforms the codeword sequence $ {b}^l$ into the reconstructed syntactic sequence $\hat{u}^m$. Only consistency between the semantic sequences $\tilde{u}^m$ and $\hat{\tilde{u}}^m$ need to be guaranteed in the coding procedures, thus the constraints between the syntactic sequences ${u}^m$ and $\hat{u}^m$ can be relaxed, which makes the coding a lossy source coding from the syntactic perspective.

\subsection{Synonymous Mappings-based Theoretical Limits}

As remarked in \cite{niu2024Mathematical}, synonymy is the critical source of relationships between the semantic information and the syntactic information since, in most instances, single-meaning semantic information has myriad presentation forms of syntactic data. Therefore, the mapping $f\left(\cdot\right)$ in \eqref{system_model} is essentially a group of synonymous mappings that map the semantic elements into different syntactic forms with the same meanings.

\begin{figure}[t]
\centering
\includegraphics[width=0.48\textwidth]{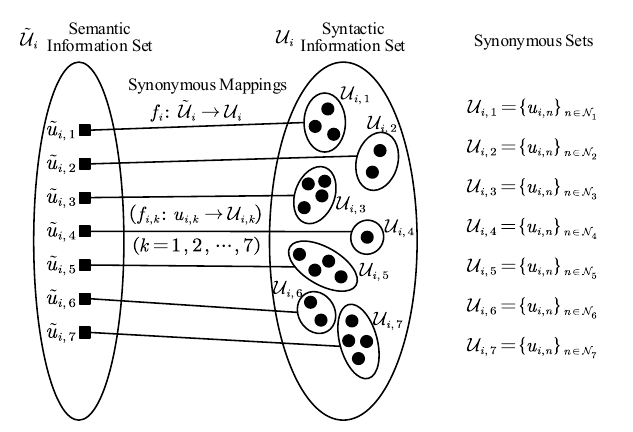}
\caption{An example of the synonymous mappings and the corresponding synonymous sets.}
\label{fig1}
\end{figure}

Figure \ref{fig1} shows an example of the synonymous mappings $f_i: \mathcal{\tilde{U}}_i \xrightarrow{} \mathcal{U}_i$ between the semantic information set $\mathcal{\tilde{U}}_i$ and the syntactic information set $\mathcal{U}_i$ for the $i$-th variable in the source sequences $\tilde{u}^m$ and ${u}^m$. From this example, a general rule can be observed: semantic elements can be mapped to an equal number of synonymous sets that represent different meanings, respectively, and contain all the possible syntactic values without overlapping between any two synonymous sets.


For an i.i.d semantic sequence $\tilde{u}$ with unified $f:\mathcal{\tilde{U}}\xrightarrow{}\mathcal{U}$ for $\forall i = 1,2,\cdots,m$, the semantic entropy $H_s\left(\mathcal{\tilde{U}}\right)$ can be expressed as
\begin{equation}
    H_s\left(\mathcal{\tilde{U}}\right) = -\sum_{k}{p\left(\mathcal{U}_k\right) \log p\left(\mathcal{U}_k\right)},
\end{equation}
where the probability of the $k$-th synonymous set
\begin{equation}
    p\left(\mathcal{U}_k\right) = \sum_{n \in \mathcal{N}_k} p\left(u_{n}\right),
\end{equation}
in which $\mathcal{N}_k$ denotes a set that contains the indexes of the syntactic values with the same meaning as the semantic element $k$.

In \cite{niu2024Mathematical}, we demonstrate that for single-symbol semantic source coding, with the semantic prefix code performed over the synonymous sets, the average code length can approach the theoretical semantic entropy limit $H_s\left(\mathcal{\tilde{U}}\right)$ without semantic losses by providing a theorem based on semantic Kraft inequality. Naturally, the same effect can be achieved by performing semantic prefix coding on the sequences, which leads to our proposed semantic arithmetic coding.

\section{Semantic Arithmetic Codes}

Consider a sequence compression procedure with arithmetic codes for the syntactic sequence $ {u}^m$, in which each syntactic variable $u_{i}$ exhibits a similar synonymous relationship like Fig. \ref{fig1}. Traditional arithmetic coding directly performs the coding procedure on each syntactic variable $u_{i}$ without considering the implicit meaning, thereby lacking certain compression efficiency for the only requirement of semantic lossless. In this section, we propose semantic arithmetic coding (SAC) using synonymous mappings for efficient semantic compression with intuitive semantic synonymy.

\subsection{The Encoding Procedure}

\begin{figure}[t]
\centering
\includegraphics[width=0.48\textwidth]{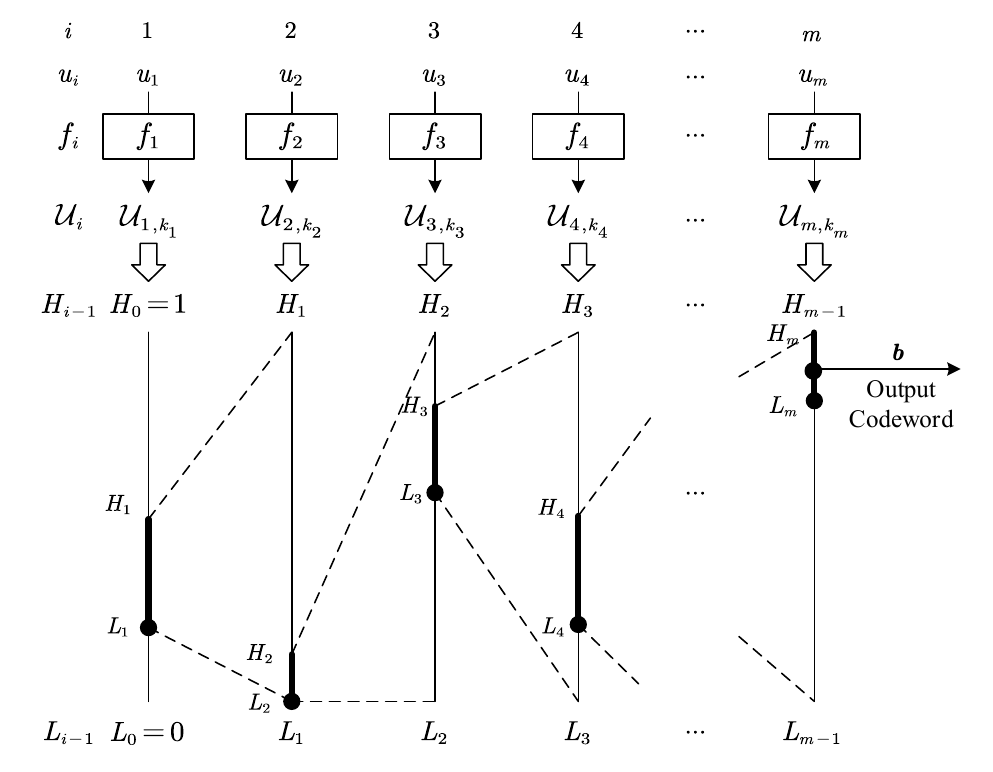}
\caption{A schematic diagram of the SAC encoding procedure.}
\label{fig2}
\end{figure}

Figure \ref{fig2} shows a general framework of the SAC encoding procedure. Similar to the traditional method \cite{cover1999elements,witten1987arithmetic}, the SAC encoder uniquely maps the message to a sub-interval on the $\left[0, 1\right)$ interval and outputs the shortest codeword represented by a binary fraction $ {b}$ in this sub-interval as the encoding result. The difference is that, to achieve semantic-oriented compression, the SAC encoder constructs synonymous mappings to partition synonymous sets for each syntactic variable and performs the coding interval update procedure over the synonymous sets.

As a specific description of the processing process in Fig. \ref{fig2}, the encoding process of the SAC encoder is detailed in Algorithm \ref{Algorithm1}. With an initialized encoding interval $\left[L_0, H_0\right) = \left[0, 1\right)$ and its interval length $R_0 = 1$, the SAC encoder performs semantic compression through a series of iterations, in which each iteration $i$ includes:

\begin{itemize}
    \item \emph{Construct synonymous mappings:} For all the syntactic values $\left\{u_{i,n}\right\}$ of the $i$-th variable, construct synonymous mappings $f_i: \mathcal{\tilde{U}}_i \xrightarrow{} \mathcal{U}_i$ to partition synonymous sets $\left\{\mathcal{U}_{i,k}\right\}_{k=1,\cdots, {|\mathcal{\tilde{U}}_i|}}$, in which $\mathcal{U}_{i,k} = \left\{u_{i,n}\right\}_{n\in\mathcal{N}_k}$;
    \item \emph{Determine synonymous set:} According to the actual value of the syntactic variable $u_i$, determine a synonymous set $\mathcal{U}_{i,r}$, such that $u_i \in \mathcal{U}_{i,r_i}, r_i \in \left\{k\right\}_{k=1,\cdots, {|\mathcal{\tilde{U}}_i|}}$;
    \item \emph{Calculate probabilities:} For all synonymous sets $\left\{\mathcal{U}_{i,k}\right\}$, calculate their probabilities with
        \begin{equation}\label{probSynoSet}
            p\left(\mathcal{U}_{i,k}\right) = \sum_{n \in \mathcal{N}_k} p\left(u_{i,k}\right).
        \end{equation}
    \item \emph{Update encoding interval:} According to the determined synonymous set $\mathcal{U}_{i,r_i}$ and the probabilities for all synonymous sets $p\left(\mathcal{U}_{i,k}\right), k=1,\cdots, {|\mathcal{\tilde{U}}_i|}$, update the encoding interval $\left[L_i, H_i\right)$ and its interval length $R_i$ with
    \begin{equation}\label{updateLHR}
        \left\{
        \begin{array}{cc}
            L_i = L_{i-1} + \sum_{k=1}^{r_i - 1} p\left(\mathcal{U}_{i, k}\right), &  \\
            H_i = L_{i} + p\left(\mathcal{U}_{i, r_i}\right) \cdot R_{i-1}, & \\
            R_i = p\left(\mathcal{U}_{i, r_i}\right) \cdot R_{i-1} = H_i - L_i.
        \end{array}
        \right.
    \end{equation}
\end{itemize}

Once the encoding interval update process corresponding to the last variable $u_m$ is completed, the SAC encoder concludes its iterations. Then it determines a shortest binary fraction $ {b}^l = \left(b_1, \cdots, b_l\right)$ as the output codeword, such that its corresponding decimal fraction $c$ belongs to the final interval $\left[L_m, H_m\right)$ and satisfies
\begin{equation}
    c = b_1 \cdot 2^{-1} + b_2 \cdot 2^{-1} + \cdots + b_l \cdot 2^{-l}.
\end{equation}

Finally, the SAC encoder transmits the output codeword $ {b}^l$ to the receiver for reconstructing the syntactic sequence. As the necessary information, the length $m$ of the syntactic sequence $ {u}^m$, along with the synonymous sets partitions of each syntactic variable and their corresponding probability information, need to be synchronized at the receiving end.

\begin{algorithm}\label{Algorithm1}
\caption{Encoding Algorithm of SAC Encoder}
\KwData{Input syntactic sequence $ {u}^m = \left(u_1, \cdots, u_m\right)$.}
\KwResult{Output codeword $ {b}$.}
\textbf{Initialize:} The encoding interval $\left[L_0, H_0\right) = \left[0, 1\right)$, interval length $R_0 = 1$, variable index $i=1$.

\Repeat{$i \le m$}{
    Construct synonymous mappings $f_i: \mathcal{\tilde{U}}_i \xrightarrow{} \mathcal{U}_i$. \\
    Determine synonymous set $\mathcal{U}_{i,r_i}$, s.t., $u_i \in \mathcal{U}_{i,r_i}$. \\
    Calculate probabilities for all synonymous sets $p\left(\mathcal{U}_{i,k}\right)$ with Eq. \eqref{probSynoSet}. \\
    Update encoding interval based on $\mathcal{U}_{i,r_i}$: $\left[L_{i-1}, H_{i-1}\right) \xrightarrow{} \left[L_{i}, H_{i}\right)$ with Eq. \eqref{updateLHR}.
}

Determine a shortest binary fraction $ {b}^l$, s.t. its corresponding decimal fraction $c \in \left[L_{i}, H_{i}\right)$.

\Return{The output codeword $ {b}^l$.}
\end{algorithm}

When only the i.i.d assumption is considered, the possible values of each syntactic variable, the synonymous mappings along with the synonymous sets partitions, and the corresponding probabilities will be exactly the same. In this case, the process of SAC encoding is equivalent to a simplified procedure that first converts the syntactic sequence into a synonymous set sequence using the predefined synonymous mapping rules, then treats the synonymous subset sequence as a new syntactic sequence, and compresses it using traditional arithmetic coding.

\subsection{The Decoding Procedure}

As a duality process of SAC encoding, the decoding procedure of the SAC decoder is presented in Algorithm \ref{Algorithm2}. To reconstruct the syntactic sequence from the received codeword $ {b}^l$, the SAC decoder initializes a decoding interval $\left[L'_0, H'_0\right) = \left[0, 1\right)$ and its interval length $R'_0 = 1$, and then utilize the decimal fraction $c$ corresponding to the binary fraction $ {b}^l$ to determine the values of each syntactic variable. The determining process is also performed through a series of iterations, in which each iteration $i$ includes:

\begin{itemize}
    \item \emph{Construct synonymous mappings:} For all the syntactic values $\left\{\hat{u}_{i,n}\right\}$ of the $i$-th reconstructed variable, construct opposite synonymous mappings $g_i: \mathcal{\hat{U}}_i \xrightarrow{} \mathcal{\hat{\tilde{U}}}_i$ according to the synchronized synonymous sets information at the sending end, and accordingly partition synonymous sets $\left\{\mathcal{\hat{U}}_{i,k}\right\}_{k=1,\cdots, {|\mathcal{\tilde{U}}_i|}}$, in which $\mathcal{\hat{U}}_{i,k} = \left\{\hat{u}_{i,n}\right\}_{n\in\hat{\mathcal{N}}_k}$;
    \item \emph{Synchronize probabilities:} To guarantee successful semantic decoding, the probabilities of all synonymous sets should be synchronized with the sending end, i.e., $p\left(\hat{\mathcal{U}}_{i,k}\right) = p\left(\mathcal{U}_{i,k}\right)$. The probability of each syntactic value needs to be assigned satisfying
    \begin{equation}
        p\left(\hat{\mathcal{U}}_{i,k}\right) = \sum_{n \in \hat{\mathcal{N}}_k} p\left(\hat{u}_{i,k}\right).
    \end{equation}
    \item \emph{Determine synonymous set:} According to the decimal fraction $c$ and the decoding interval $\left[L'_{i-1}, H'_{i-1}\right)$, determine the reconstructed synonymous set $\hat{\mathcal{U}}_{i,r_i},  r_i \in \left\{k\right\}_{k=1,\cdots, {|\mathcal{\tilde{U}}_i|}}$,  such that satisfying
    \begin{equation}\label{decideSynoSet}
        \frac{c - L'_{i - 1}}{R'_{i-1}} \in \left[\sum_{k=1}^{r_i - 1} p\left(\hat{\mathcal{U}}_{i, k}\right), \sum_{k=1}^{r_i } p\left(\hat{\mathcal{U}}_{i, k}\right)\right).
    \end{equation}
    \item \emph{Export syntactic value:} Select a syntactic value $\hat{u}_{i,n}$ from the determined synonymous set $\hat{\mathcal{U}}_{i,r_i}$ as the reconstructed syntactic value $\hat{u}_{i}$. It can be randomly chosen based on the normalized probability of each syntactic value in the determined synonymous set
    \begin{equation}
        p'\left(\hat{u}_{i, n}\right) = \frac{p\left(\hat{u}_{i, n}\right)}{p\left(\hat{\mathcal{U}}_{i, r_i}\right)}, n\in\mathcal{N}_{r_i},
    \end{equation}
    or guided by specific semantic background knowledge related to the semantic source.
    \item \emph{Update decoding interval:} According to the determined synonymous set $\hat{\mathcal{U}}_{i,r_i}$ and the probabilities for all synonymous sets $p\left(\mathcal{\hat{U}}_{i,k}\right), k=1,\cdots, {|\mathcal{\tilde{U}}_i|}$, update the decoding interval $\left[L'_i, H'_i\right)$ and its interval length $R'_i$ with
    \begin{equation}\label{updateLHR_D}
        \left\{
        \begin{array}{cc}
            L'_i = L'_{i-1} + \sum_{k=1}^{r_i - 1} p\left(\hat{\mathcal{U}}_{i, k}\right), &  \\
            H'_i = L'_{i} + p\left(\hat{\mathcal{U}}_{i, r_i}\right) \cdot R'_{i-1}, & \\
            R'_i = p\left(\hat{\mathcal{U}}_{i, r_i}\right) \cdot R'_{i-1} = H'_i - L'_i.
        \end{array}
        \right.
    \end{equation}
\end{itemize}

Once the decoding interval update process corresponding to the last reconstructed variable $\hat{u}_m$ is completed, the SAC decoder concludes its iterations, and outputs a combination of reconstructed syntactic values $\hat{u}^m = \left(\hat{u}_1, \cdots, \hat{u}_m\right)$ as the reconstructed syntactic sequence.

Same as the SAC encoder, when only the i.i.d assumption is considered, the synonymous mappings along with the synonymous sets partitions, and the corresponding probabilities will be exactly the same. In this case,  the process of SAC decoding is equivalent to a simplified procedure that first reconstructs the synonymous set sequence with the traditional arithmetic decoder, and determines each syntactic variable based on the unified synonymous set partition rules.

To summarize, SAC implements semantic compression and reconstruction by constructing reasonable synonymous mappings and performing arithmetic coding procedures over synonymous sets.

\begin{algorithm}\label{Algorithm2}
\caption{Decoding Algorithm of SAC Decoder}
\KwData{Received codeword $ {b}^l$, sequence length $m$.}
\KwResult{Output reconstructed syntactic sequence $\hat{u}^m$.}
\textbf{Initialize:} The decoding interval $\left[L'_0, H'_0\right) = \left[0, 1\right)$, interval length $R'_0 = 1$, variable index $i=1$.

Recover the decimal fraction $c$ corresponding to the codeword $ {b}^l$.

\Repeat{$i \le m$}{
    Construct synonymous mappings $g_i: \mathcal{\hat{U}}_i \xrightarrow{} \mathcal{\hat{\tilde{U}}}_i$. \\
    Synchronize probabilities for all synonymous sets $p\left(\hat{\mathcal{U}}_{i,k}\right) = p\left(\mathcal{U}_{i,k}\right)$. \\
    Determine synonymous set $\hat{\mathcal{U}}_{i,r_i} \in \left\{\mathcal{U}_{i,k}\right\}$, s.t., satisfying Eq. \eqref{decideSynoSet}. \\
    Export syntactic value $\hat{u}_i$, s.t., $\hat{u}_i \in {\hat{\mathcal{U}}_{i,k_i}}$. \\
    Update decoding interval based on $\hat{\mathcal{U}}_{i,r_i}$: $\left[L'_{i-1}, H'_{i-1}\right) \xrightarrow{} \left[L'_{i}, H'_{i}\right)$ with Eq. \eqref{updateLHR_D}.
}

Combine output sequence $ {\hat{u}}^m = \left(\hat{u}_1, \cdots, \hat{u}_m\right)$.

\Return{The output syntactic sequence $ {\hat{u}}^m$.}
\end{algorithm}

\subsection{Theoretical Limits Analysis}

Herein, we analyze the theoretical semantic compression limits of our proposed SAC, based on the extension of the code length theorem of the classical arithmetic coding algorithm \cite{cover1999elements} to the following semantic version:

\newtheorem{theorem}{Theorem}
\begin{theorem}
    For a semantic arithmetic coding procedure, given any syntactic sequence $ {u}^m$ with the probability mass function of its corresponding synonymous set sequence $q\left(\mathcal{U}_{1,r_1},\ldots,\mathcal{U}_{m,r_m}\right)$, it enables one to encode ${u}^m$ in a code of length $-\log q\left(\mathcal{U}_{1,r_1},\ldots,\mathcal{U}_{m,r_m}\right) + 2$ bits.
\end{theorem}

This theorem can be simply proved by replacing the probability mass function for the syntactic sequence in the compression limit of arithmetic coding \cite{cover1999elements} with the probability mass function for the synonymous set sequence, in which the compression limit of arithmetic coding is fundamentally derived from the corresponding analysis of Shannon-Fano-Elias codes \cite{cover1999elements, cover1973enumerative}.

With i.i.d assumption and the assumed distribution $q$ being equal to the true distribution $p$, the average code length can approach the semantic entropy limits $H_s\left(\tilde{\mathcal{U}}\right)$ if $m \xrightarrow{} \infty$, i.e.,
\begin{equation}
     H_s\left(\tilde{\mathcal{U}}\right) < \bar{L}_s \le \lim_{m \xrightarrow{} \infty} \frac{- \log p\left(\mathcal{U}_{1,r_1},\cdots,\mathcal{U}_{m,r_m}\right) + 2}{m},
\end{equation}
in which the upper bound approaches $H_s\left(\tilde{\mathcal{U}}\right)$, thereby proving the theoretical limit approachability of our proposed SAC.

\begin{figure}
\centering
\includegraphics[width=0.48\textwidth]{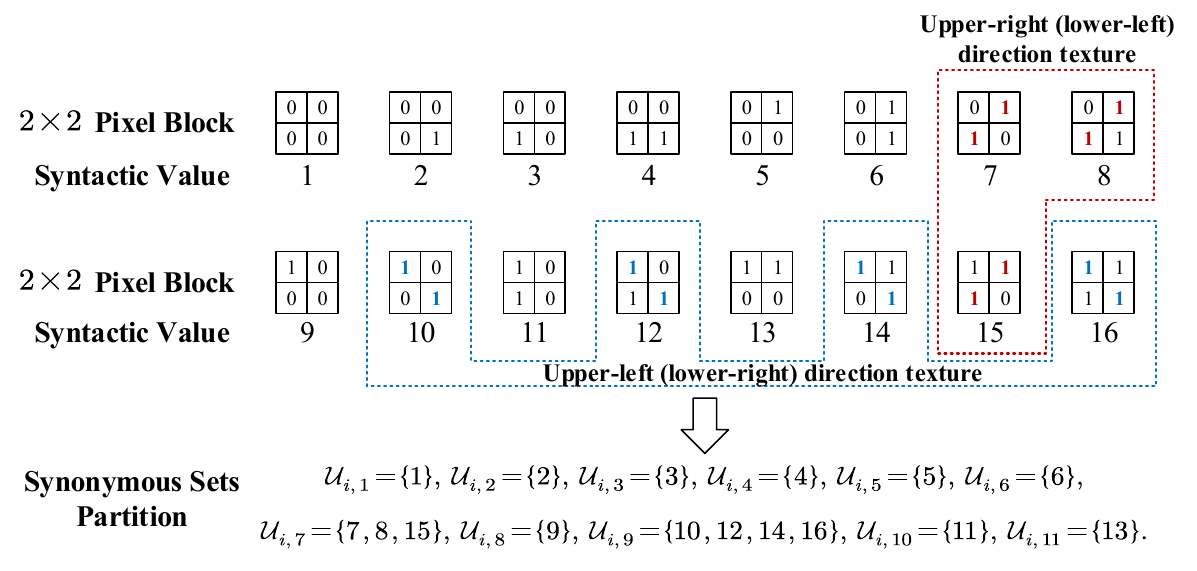}
\caption{The synonymous sets partition for edge texture maps.}
\label{fig3}
\end{figure}

\section{Experimental Results}

\begin{figure*}[t]
\centering
\includegraphics[width=0.99\textwidth]{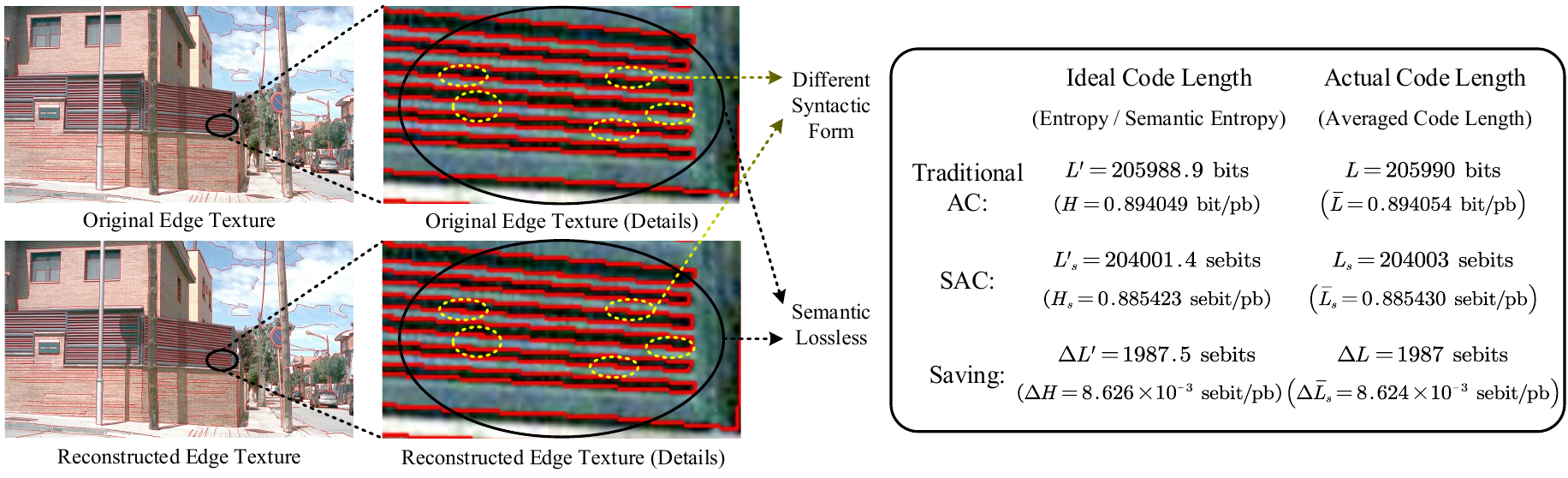}
\caption{An example of the compression and the reconstruction effects for edge texture map semantic compression, in which ``sebits'' denotes semantic bits for the resulting unit of semantic source coding, presented by \cite{niu2024Mathematical}. Besides, ``bit/pb'' and ``sebit/pb'' respectively denote bit per pixel block and sebit per pixel block, acting as the unit of the entropy and the average code length of traditional AC and our proposed SAC based on our coding configuration.  }
\label{fig4}
\end{figure*}

In this section, we verify the semantic compression performance of our proposed SAC along with its ability to preserve semantics.

We consider a scenario of semantic compression for edge texture maps of natural images, in which the edge texture maps and their corresponding natural images are all derived from the BIPEDv2 dataset \cite{soria2023dense}. We use the edge texture in natural images as a form of semantic information contained in the images, and employ the annotated edge texture maps as the syntactic source for semantic compression and reconstruction. The resolutions of the annotated edge texture maps are $1280 \times 720$, in which pixels representing objects and the background are labeled as value $0$, and those of the edge texture between different objects and between an object and the background are labeled as value $1$.

To construct the synonymous mappings, we regard each non-overlapped $2\times2$ pixel block as a syntactic symbol and partition the $16$ syntactic symbols into $11$ synonymous subsets according to Fig. \ref{fig3}, in which each synonymous set represents a type of local edge texture semantic. Therefore, the edge texture maps can be considered a syntactic sequence of length $m = 640\times360 = 230400$. On this basis, we assume that syntactic sequences satisfy the i.i.d. assumption, and the probabilities of syntactic symbols and synonymous sets in the encoding and decoding process are directly obtained based on probability statistics and synchronized at both ends.

We perform our proposed SAC on the test set, consisting of 50 edge texture maps along with their corresponding natural images, to verify the semantic compression effect, and utilize the traditional AC method as the comparison scheme. As a representative result, Fig. \ref{fig4} shows an example of the compression and the reconstruction effects for edge texture maps semantic compression with our proposed SAC, in which the original and the reconstructed edge texture maps are both labeled on the corresponding natural image to verify the semantic accuracy of the edge texture semantics. From the perspective of the reconstruction effect, although the syntactic form of the reconstructed edge texture differs from the original edge texture, it does not affect the accuracy of its edge texture semantics in this natural image. This observation is consistent with the effects seen in the other samples in the test set, indicating that no semantic losses exist on the reconstructed edge texture map with our proposed method.

On this basis, compression efficiency has been improved with SAC to some extent. From the perspective of the actual code length, SAC provides an effective compression efficiency improvement compared with the traditional method. In the example shown in Fig. \ref{fig4}, SAC saves 1987 sebits \cite{niu2024Mathematical} compared with traditional arithmetic coding, equivalent to a 0.96\% improvement in compression efficiency. Additionally, from another viewpoint, the averaged code length performed by SAC can break through the Shannon entropy of the classical information theory, and further approximate the theoretical semantic compression limits, i.e., semantic entropy, with a gap of $7 \times 10^{-6}$ sebit per pixel block. As for the entire test set, SAC can save 1935.38 sebits of average code length compared to traditional methods, equivalent to a 1.36\% improvement in average compression efficiency. Furthermore, an average gap of $4 \times 10^{-6}$ sebit per pixel block between the code length of SAC and semantic entropy can be achieved on the test set.

These results effectively demonstrate the performance of our proposed SAC method, i.e., it can achieve an effective compression efficiency improvement and approximate the semantic entropy with semantic lossless.

\section{Conclusion}

In this paper, we propose a semantic source coding method called semantic arithmetic coding. By constructing reasonable synonymous and performing arithmetic coding procedures over synonymous sets, the compression efficiency can be improved compared with the traditional arithmetic codes with semantic lossless. Additionally, we provide a theoretical limit analysis of our proposed method based on an extension code length theorem of arithmetic codes, along with experimental verification, to confirm its approachability to semantic entropy.

\section*{Acknowledgment}

This work was supported by the National Natural Science Foundation of China (No. 62293481, No. 62071058).

\IEEEtriggeratref{14}

\bibliographystyle{IEEEtran}
\bibliography{IEEEabrv, ref}










\end{document}